\documentclass[12pt]{article}
\usepackage{rotating}
\usepackage{epsfig,amssymb}
\input{epsf}

\def\laq{\raise 0.4 ex \hbox{$<$}\kern -0.8 em\lower 0.62 ex\hbox{$\sim$}}
\def\gaq{\raise 0.4 ex \hbox{$>$}\kern -0.7 em\lower 0.62 ex\hbox{$\sim$}}

\def\beq{\begin{equation}}
\def\eeq{\end{equation}}
\def\beqa{\begin{eqnarray}}
\def\eeqa{\end{eqnarray}}

\def\to{\rightarrow}

\def\to{\rightarrow}

\def\CQG{{\it Class. Quantum Gravity} }

\def\NP{{\it Nucl. Phys.} }
\def\PL{{\it Phys. Lett.} }

\def\PR{{\it Phys. Rev.} }
\def\PRL{{\it Phys. Rev. Lett.} }

 \def\frac#1#2{{\textstyle{{#1}\over {#2}}}}
 \def\lsim{\mathrel{\rlap{\lower4pt\hbox{\hskip1pt$\sim$}}
    \raise1pt\hbox{$<$}}} \def\gsim{\mathrel{\rlap{\lower4pt\hbox{\hskip1pt$\sim$}}
    \raise1pt\hbox{$>$}}}
\def\sqr#1#2{{\vcenter{\vbox{\hrule height.#2pt
         \hbox{\vrule width.#2pt height#1pt \kern#1pt
         \vrule width.#2pt}
         \hrule height.#2pt}}}}

\def\gappeq{\mathrel{\rlap {\raise.5ex\hbox{$>$}} {\lower.5ex\hbox{$\sim$}}}}

\def\lappeq{\mathrel{\rlap{\raise.5ex\hbox{$<$}}
{\lower.5ex\hbox{$\sim$}}}}

\begin{document}
\pagestyle{plain}

\vspace*{1.0cm}

\begin{flushright}
DF/IST--4.2009\\
March 2009
\end{flushright}
\vspace{15mm}

\begin{center}

{\Large\bf The cosmological constant problem: a user's guide$^{*}$}

\vspace*{1.0cm}

Orfeu Bertolami$^{**}$ \\
\vspace*{0.5cm}
{Instituto Superior T\'ecnico, Departamento de F\'\i sica, \\
Av. Rovisco Pais, 1049-001 Lisboa, Portugal}\\

\vspace*{2.0cm}
\end{center}

\begin{abstract}

\noindent
We discuss the validity of general relativity at low-energy and relate the threshold below which
the theory breaks down with the observed value of the cosmological constant. This suggests the existence of a 
mass scale of ${\cal O}(10^{-3})~eV$ and a putative violation of the equivalence principle at about $10^{-14}$ level.

\end{abstract}

\vspace*{2.0cm}

\begin{center}

{\bf Dedicated to the memory of Maria da Concei\c c\~ao Bento}

\end{center}

\vfill
\noindent\underline{\hskip 140pt}\\[4pt]
{$^*$ Essay selected for an honorable mention by the Gravity Research Foundation, 2009} \\
\noindent
{$^{**}$ Also at Instituto de Plasmas e Fus\~ao Nuclear, Instituto Superior T\'ecnico,
Lisboa} \\
{E-mail address: orfeu@cosmos.ist.utl.pt}

\newpage
\section{Introduction}
\label{sec:intro}

Einstein's theory of
general relativity as established by its field equations is
consistent with all experimental evidence to considerable accuracy
(see e.g. Refs. \cite{Will05,BPT06} for
reviews). However, quite fundamental problems lead one to conclude that
general relativity cannot be a complete description of gravity.
These difficulties are associated with
the existence of spacetime singularities, the
cosmological constant problem and the incompatibility of the
theory with quantum mechanics. Indeed, all attempts to quantize
gravity with the techniques of quantum gauge field theories that successfully
describe electromagnetic, weak, and strong interactions, were shown to be unfeasible, as
beyond one-loop (without matter), the quantum theory emerging from
general relativity is not renormalizable. Furthermore, none of the
known approaches to quantize gravity do solve these difficulties. Indeed, for instance, in
the context of superstring/M-theory,
the most studied programme to quantize gravity, singular solutions such as black
holes are encountered, and no satisfactory answer to the cosmological constant problem has ever been presented \cite{Witten}.

Furthermore, the curvature-matter connection of Einstein's theory is not the most general one and,
having in mind the Standard Model (SM) of the fundamental interactions and the cosmological standard model, 
which requires an early period of accelerated expansion, i.e. inflation,
it is fairly natural to consider additional fields, most particularly
scalar fields. In this respect, scalar-tensor theories of gravity are an interesting possibility
given that they capture the main features of many unification models. For instance,
the  graviton-dilaton system that arises from string/M-theory can be
seen as a particular example of a scalar-tensor theory of gravity.

However, likewise general relativity, all proposed alternative gravity theories do not provide
a consistent description of
our universe given the huge discrepancy between the observed value of the cosmological constant
and the one arising from the SM. Several proposals have been advanced to solve this difficulty
(see e.g. Refs. \cite{Weinberg,Carroll,Bauer}). A vanishing cosmological constant suggests a symmetry principle in action.
Indeed, it has been suggested that the problem admits a solution likewise
the strong CP problem \cite{Wilczek}, which actually can be somehow implemented in the
context of a S-modular invariant $N=1$ supergravity quantum cosmological model in
a closed homogeneous and isotropic spacetime \cite{BSchiappa}. A connection between the cosmological constant
and Lorentz invariance has also been discussed in different contexts \cite{Bertolami97, BC06}
as well as an invariance under complex transformations \cite{Hooft06,Andrianov}. Finally,
an evolving cosmological term has also been considered by some authors
\cite{Bronstein,Bertolami86,Ozer}.
Unfortunately, none of the proposed mechanisms are fully consistent
\cite{Weinberg,Carroll,Bauer}.

As already mentioned, within the framework of supertring/M-theory,
no fully satisfactory solution for the cosmological constant problem has ever been advanced \cite{Witten}, even though more
recently, it has been argued that a solution arises through a suitable choice of the vacuum
in the ``landscape'' of vacua of the theory in its
multiverse interpretation (see e.g. \cite{ST06} and references therein). In this context, each vacuum configuration in
the multitude of about $10^{500}$ vacua of the theory \cite{BoussoPol} is regarded as a distinct universe,
a ``bubble'' in an {\it a priori} spacetime.
The evolution of each universe follows classical and quantum physics with physical parameters and coupling constants corresponding to each vacuum state.
The choice of a suitable vacuum for
our universe requires some selection criteria.
Anthropic arguments \cite{Susskind,Polchinski} and quantum cosmological considerations \cite{HMersini} have
been proposed as possible vacuum selection procedures.

Another interesting possibility is that these
``bubble'' universes do interact with each other
as it usually occurs in condensate matter systems with multiple vacua.
An interaction between universes,
based on a Curvature Principle, has been suggested and shown,
in the context of a simplified model, that it can drive
the cosmological constant of one of the universes toward a vanishingly small
value \cite{Bertolami08}. It has also been conjectured on
how this Curvature Principle suggests a solution for the entropy
evolution of the
universe so to satisfy the generalized Second principle of Thermodynamics \cite{Bertolami08}.

In this work one considers the cosmological constant problem from a quite conservative standing point, i.e. from the point of view
of general relativity itself and of the SM, regardless of any further
unification of the fundamental interactions. It is argued that although
the cosmological constant problem is an ultraviolet problem,
possibly involving non-local aspects, and its solution cannot
be found in the context of general relativity,
it is nevertheless, interesting to examine what can be learnt by assessing the
validity of the theory at low energies.
These considerations lead one to
conclude that although the fine tuning related with ultraviolet behaviour of the theory
involves a still eluding mechanism, the cosmological constant problem has also a bearing at low energy, and, in particular, it indicates the existence a
mass scale of about ${\cal O}(10^{-3})~eV$ and a putative violation of the equivalence principle at $10^{-14}$ level so to account 
for the residual non-vanishing value of the vacuum energy.

\section{Relativistic invariance, four-dimensional cutoff and a fundamental scalar field}
\label{sec:model}

In what follows one considers the vacuum energy contribution arising
from a field theory calculation of a new fundamental scalar field with mass $m$.
But before one addresses this issue it is relevant to point out that the vacuum energy is an
extensive quantity and hence it is an additive part of the energy of a physical system.
This energy is irrelevant in non-gravitational physics where only energy differences matter;
however, the situation is fundamentally different in general relativity given that
in its framework any non-vanishing
configuration of energy/matter is a source for gravity and curves spacetime.
Actually, strictly speaking, the vacuum energy shows up in both sides
of Einstein's field equations. Indeed, it reveals itself as
a constant, $\Lambda_E$, at the geometrical side of the field
equations and as a source term from the vacuum contribution of the relevant
fields:
\beq
G_{ab} + \Lambda_E g_{ab} = 8 \pi G (T_{ab} + <T_{ab}>)~~,
\label{eq:einstein}
\eeq
where $G_{ab}$ is the Einstein tensor and $<T_{ab}>$ is the contribution to
vacuum of the field(s) after, for instance, the matter theory undergoes some
spontaneous symmetry breaking. The cosmological constant problem consists in
the highly unnatural fine tuning between $\Lambda_E$ and $8 \pi G<T_{ab}>$
so to fit the cosmological data.
Furthermore, given Lorentz invariance, then $<T_{ab}> = \rho_V g_{ab}$, where
$\rho_V$ is the vacuum energy density. The vacuum pressure corresponds to the space
components $<T_{ii}>$, from which follows that energy density and pressure
must satisfy the relationship:
\beq
\rho_V = - p_V~~.
\label{eq:eqstate}
\eeq
Assuming the observed cosmological value for the vacuum energy,
namely, $\Omega_{\Lambda}=0.7$, or
$\rho_V= 5.65 \times 10^{-47}~GeV^4$ for $h_0=0.7$\footnote{One should remark that
although cosmological data do not exclude other scenarios, supernovae data, baryon acoustic
oscillations, microwave background radiation shift parameter and topological considerations are
all consistent with the possibility that the late accelerated expansion of the universe
is driven by the cosmological constant
(see for instance \cite{Data,DEMG06} and references therein).}, comparison with
the contribution from the SM after the Higgs field acquires a non-vanishing vacuum
expectation value, $\rho_V^{SM} = {\cal O}(250~GeV)^4$, yields a discrepancy of
$10^{56}$ orders of magnitude or an adjustment of $56$ decimal places.

In what concerns the field theory computation of the energy density in the
context of a real scalar field theory, it has been recently pointed out that only through a
fully covariant $4$-dimensional procedure one can respect the
vacuum equation of state, Eq. (\ref{eq:eqstate})
\cite{Akhmedov}. Indeed, the zero-point energy of a free real scalar field $\phi$ of mass
$m$ can be estimated introducing into the energy-momentum tensor $T_{ab}$ of the field
a plane-wave decomposition and taking the vacuum expectation
value of the component $T_{00}$. It then follows the well-known
result for the contribution of the field $\phi$ to the energy density of
the vacuum:
\beq
\rho_V={1 \over 2}\int_0^{\Lambda} {d^3k \over (2\pi)^3}\sqrt{{\bf k}^2+m^2}~~,
\label{rho1}
\eeq
for a high-energy cutoff $\Lambda$. Notice that this is just the
sum over all modes of the zero-point energies
$\omega_{k}/2=\sqrt{{\bf k}^2+m^2}/2$ and
that integral in Eq. (\ref{rho1}) is clearly divergent as $\Lambda \to \infty$.

The contribution to the vacuum pressure can be obtained through the same procedure:
\beq
p_V={1 \over 6}\int_0^{\Lambda}{d^3k \over (2\pi)^3}{{\bf k}^2 \over \sqrt{{\bf k}^2+m^2}}~~,
\label{press1}
\eeq
where the isotropy of the vacuum has been used. The above computations were carried
out in Minkowski space, but
it can be argued that the spacetime curvature does not affect results Eqs.
(\ref{rho1}) and (\ref{press1}) \cite{Akhmedov}.

In order to implement a relativistically
invariant 4-momentum cutoff, one must use manifestly covariant expressions
for $\rho_V$ and $p_V$. Thus, one must turn the integrals Eqs.
(\ref{rho1}) and (\ref{press1}) over $3$-momentum
into $4$-momentum integrals. This is not a well defined procedure given that the integrals
are divergent and therefore, one must regularize them so not to modify the
integrands and not to spoil their covariance properties. A suitable procedure involves
differentiating the integrals with respect to $m^2$.
From Eq. (\ref{rho1}) one obtains
\beq
{\partial \rho_V \over \partial m^2}={1 \over 2}\int_0^{\Lambda}{d^3k \over (2\pi)^3
2\omega_{k}}~~.
\label{drho1}
\eeq
The r.h.s. integral is Lorentz
invariant and can be written in a manifestly invariant form as \cite{Akhmedov}:
\beq
{\partial \rho_V \over \partial m^2}={i \over 2}\int_0^{\Lambda}{d^4k \over (2\pi)^4}
{1 \over k^2-m^2+i\varepsilon}~~.
\label{drho2}
\eeq
Integrating Eq. (\ref{drho2}) one gets, up to an $m$-independent
constant,
\beq
\rho_V=
{1 \over 2}\int_0^{\Lambda}{d^4 k_E \over (2\pi)^4}\ln\left(1+{m^2 \over k_E^2}\right)~~,
\label{rho3}
\eeq
where the integral was Wick-rotated to Euclidean space, so that a
4-momentum cutoff can be introduced. With the now covariant cutoff $\Lambda$ one finds from Eq. (\ref{rho3}) \cite{Akhmedov}
\beq
\rho_V={1 \over 64\pi^2}\left[\Lambda^4\ln\left({\Lambda^2+m^2 \over \Lambda^2}
\right)+\Lambda^2 m^2
-m^4\ln\left({\Lambda^2+m^2 \over m^2}\right)\right]~~.
\label{rho4}
\eeq
This result accounts for the vacuum contribution $<T_{ab}>$ in the r.h.s. of Einstein's equations.
Restricting oneself to the SM, hence the cutoff, $\Lambda={\cal O}(250~GeV)$.

In what concerns the geometrical side of Einstein's equations in order to ensure the proper cancelation of the
huge vacuum contribution arising from the SM one should consider, up to the $8 \pi G$ factor
and the still unknown cancelation mechanism, the estimate Eq. (\ref{rho3}).
However, since it may happen that Einstein's equations break down at
low energies, one considers instead the integration from
a low energy cutoff $\Delta$ to $\Lambda$:
\beq
\Lambda_E=4 \pi G\int_{\Delta}^{\Lambda}{d^4 k_E \over (2\pi)^4}\ln\left(1+{m^2 \over k_E^2}\right)~~,
\label{lambda1}
\eeq
and hence the resulting cosmological term in Einstein's equation is the ``regularized'' one:
\beq
\Lambda_{Reg.}=-{G \over 8 \pi}\left[\Delta^4\ln\left({\Delta^2+m^2 \over \Delta^2}
\right)+\Delta^2 m^2
-m^4\ln\left({\Delta^2+m^2 \over m^2}\right)\right]~~.
\label{lambda2}
\eeq
Notice that the negative sign reflects the fact that this contribution is in the geometrical side of
Einstein's field equation, but it clearly corresponds to a residual positive vacuum energy density.

In what follows one assumes that the scalar field mass is of the order of the low-energy cutoff and therefore
\beq
\Lambda_{Reg.}=-{G \over 8 \pi}\Delta^4~~.
\label{lambda3}
\eeq
The scale $\Delta$ can be estimated by comparison with the observed value of the
vacuum energy density. It then follows that $\Delta={\cal O}(10^{-3})~eV$ and hence
\beq
{\rho_{Reg.} \over \rho_{SM}}\simeq 10^{-56}\equiv \delta^4~~.
\label{lambda4}
\eeq
Thus, the scale $\delta=10^{-14}$ corresponds, relative to the SM typical energy scale, the
scale below which general relativity breaks down. The most likely broken low energy symmetry to
consider is the {\it equivalence principle}
which states that {\it any energy couples to gravity in the same fashion}.
This is fairly logical as a violation of the equivalence principle in this instance would
mean that the vacuum energy couples to gravity in a special way and curves spacetime somewhat differently. These considerations lead one to
conclude that this coupling principle holds down to ${\cal O}(10^{-3})~eV$.
It is relevant to point out that
the current bound on the validity of the equivalence principle is $1.4 \times 10^{-13}$ \cite{Adelberger}.

Notice that this conjectured violation of the equivalence principle is, at least at first glance,
unrelated with the recently reported one encountered at cluster and cosmological scales, which are due to dark matter-dark matter interaction \cite{Kesden} or to the interaction of dark matter
to dark energy \cite{Bertolami07a,Bertolami07b}.
However, given that dark energy might be the
vacuum energy density, one can thus speculate that these pieces of evidence
about the violation of the
equivalence principle might have a common origin. If so, then this violation should
arise at the typical ``length scale'' of dark energy or vacuum energy \cite{Perl}:
\beq
L_{DE}= \left[h c \over \rho_V \right]^{1/4}=8.5 \times 10^{-5}~m = 85 \mu m~~,
\label{length}
\eeq
where one expects Yukawa-type deviations from Newton's gravitation potential.
No deviations have been found that are at least as important
as gravity for distances of about $56~\mu m$ \cite{Kapner}. For smaller distances,
no conclusions can be drawn as contributions significantly greater than gravity are
still consistent with the experimental results, but they cannot be disentangled from
microscopical forces of electrostatic nature and the Casimir effect \cite{Geraci}. Of course, our conclusions suggest however, that
deviations should be found corresponding to violations of the equivalence
principle of $10^{-14}$
and at separation distances no very different than (\ref{length}).

It is interesting to point out that similar conclusions have been
drawn by arguing that the vacuum energy
has its origin in a scalar field \cite{Beane} or in
tensor fields through the spontaneous violation of Lorentz
invariance \cite{Bertolami97}.

\section{Discussion and Conclusions}

The cosmological constant problem challenges the current understanding about the vacuum of
the SM and its relationship with gravity. In the context of general relativity,
the cosmological constant that appears in the geometrical side of the Einstein's
field equations must be employed to cancel out
the contribution arising from the contribution to the vacuum energy of the matter fields.
In the SM, the Higgs
field contribution yields a discrepancy of $56$ orders of magnitude
with the observed value of the
vacuum energy on cosmological scales. This requires an adjustment
between the two quantities in Einstein's equation of $56$ decimal places. Of course, further
unifying schemes designed to encompass the strong interaction,
and eventually gravity, do turn this tuning much tighter.
It is believed that only in the context of a
theory of quantum gravity this adjusting problem can be properly addressed.
From this point of view, the cosmological constant problem is a touching stone as
any theory that does not satisfactorily solve this riddle cannot be regarded
as fundamental.

In this work it has been considered the possibility that the besides
its well known ultraviolet inadequacy,
general relativity may turn out to be unsuitable at low energies and
that this may account for the residual value of the vacuum energy.
After remarking that only through the implementation
of Lorentz invariant 4-dimensional cutoff one can satisfy
the invariance of the vacuum under Lorentz transformations,
one has shown, in the context of a field theory of a real scalar field
with mass $m$, that if the low-energy cutoff
$\Delta$ is equal to $m$, than one should expect a violation
of the equivalence principle at the level of the
observed vacuum energy density, that is at about ${\cal O}(10^{-12})~eV^4$.
The typical length scale associated
with this energy is about $85~\mu m$. This violation might appear as
Yukawa-type deviations from the Newtonian
gravitational potential.

Searching for this violation is therefore of great relevance. However,
confirmation of the proposed scenario
requires the detection of the fundamental scalar
field\footnote{Notice that a long lived fundamental scalar field has also been proposed as
a candidate for dark matter \cite{Nunes99,Bento01}.}. In this respect,
it is particularly interesting to conjecture
that this scalar field corresponds to the quantum
excitation of the cosmological quintessence field responsible
for the late time accelerated expansion of the universe. This is a quite
exciting possibility as it has been argued elsewhere \cite{BRosenfeld08}
that these excitations could, under conditions, be detected at LHC or
at the next generation of colliders \cite{BRosenfeld08}.

\vspace{0.5cm}


\noindent



\bibliographystyle{unstr}

\end{document}